# Coupling of a whispering gallery mode to a silicon chip with photonic crystal

YUYANG ZHUANG,[1,2] HAJIME KUMAZAKI,[1] SHUN FUJII,[1] RIKU IMAMURA,[1] NURUL ASHIKIN BINTI DAUD,[1] RAMMARU ISHIDA,[1] HEMING CHEN,[2] TAKASUMI TANABE[1,*]

[1]*Department of Electronics and Electrical Engineering, Faculty of Science and Technology, Keio University, Yokohama 223-8522, Japan*
[2]*Department of Electronic and Optical Engineering, Nanjing University of Posts and Telecommunications, Nanjing 210023, China*
*\*Corresponding author: takasumi@elec.keio.ac.jp*

**We demonstrate the efficient coupling (99.5%) of a silica whispering gallery mode microresonator directly with a silicon chip by using a silicon photonic crystal waveguide as a coupler. The efficient coupling is attributed to the small effective refractive index difference between the two devices. The large group index of the photonic crystal waveguide mode also contributes to the efficient coupling. A coupling $Q$ of $2.68 \times 10^6$ is obtained, which allows us to achieve the critical coupling of a silica whispering gallery mode with an intrinsic $Q$ of close to $10^7$ with a Si chip**.

Whispering gallery mode (WGM) optical microresonators with an ultra-high quality factor ($Q$) can exhibit a high intensity optical field, with which many applications have been demonstrated, including all-optical switching [1, 2], optical buffer [3], cavity QED [4, 5], Brillouin lasing [6, 7], and dynamic wavelength tuning [8]. However, the major challenge that must be met as regards the application of optical microresonators is their integration with an optical coupling mechanism.

To obtain a high coupling efficiency, phase-matching is the critical condition that must be satisfied, which means the propagation constant of the light in the coupler must be well matched to that of the mode in the optical microresonator. In other words, the effective refractive index ($n_\text{eff}$) of the waveguide and resonator modes must be close to each other.

In normal circumstances, the $n_\text{eff}$ of the fundamental propagation mode of a waveguide is close to the refractive index of the material of which the waveguide device is made. There are various kinds of devices, including tapered fibers [9], side-polished fibers [10], prisms [11], and planar waveguides [12, 13], that can couple light out from the WGM. Of these couplers, tapered fibers have become the most widely-used method thanks to their high coupling efficiency (~99.97%) [14] and easy alignment. However, tapered fiber fabrication requires skilled researchers. Moreover, they are sensitive to mechanical vibrations during the measurement. Compared with tapered fibers, prisms have significantly higher robustness, and a reasonably high coupling efficiency (~80%) [15]. However, this approach is based on free space optics and demands precise alignment when adjusting the angle of the incident beam.

Recently, the demand for the direct coupling of high-$Q$ WGMs to planar photonic circuits has been increasing for cases such as the integration and locking of III-V material lasers with a frequency comb source generated by ultrahigh-$Q$ WGM microresonators [16]. Although coupling a WGM to a chip is promising since such a setup can expand the application of WGM resonators, the biggest challenge is the large refractive index mismatch between high-$Q$ microresonators made of relatively low index materials (e.g., SiO$_2$; $n_\text{silica}$ = 1.44) and the commonly used photonic platforms made of high index materials such as silicon (Si) (e.g., Si; $n_\text{Si}$ = 3.48).

There have been some demonstrations of direct coupling that proved to be robust. However, to date the demonstrated planar waveguide couplers have a common factor linking them, namely that the refractive indices of the materials used in the microresonators and waveguides are relatively close. A MgF$_2$ crystal optical resonator has been coupled to a silica beam waveguide [17], and their refractive indices are both close to 1.44. A lithium tantalite (LT, $n_\text{LT}$ = 2.17) microresonator has been coupled to a Si waveguide, both with a relatively high refractive index [18]. However, the coupling of the WGM with the highest $Q$ (MgF$_2$ and silica) to a monolithic Si platform has yet to be demonstrated due to the large refractive index mismatch.

In this letter, we report extremely efficient coupling between a silica WGM and a Si monolithic photonic crystal (PhC) waveguide. We show that the PhC waveguide satisfies the phase-matching condition and enables efficient coupling even when the two materials have very different refractive indices.

The coupling structure is depicted in Fig. 1(a), where an edge silica toroid WGM microresonator is placed on top of a W0.98 (98% of the original width) Si PhC waveguide. The PhC slab was fabricated using a photolithographic CMOS compatible process at a silicon photonics foundry.

Figure 1(b) is a scanning electron microscope image of the fabricated W0.98 PhC waveguide, where the lattice constant, air-hole diameter and slab thickness are 420, 256, and 210 nm, respectively. The silica cladding is removed and an air-bridge structure is formed to enable the direct coupling of WGMs in the Si PhC waveguide mode. The calculated electric field distribution for this propagation mode is shown in Fig. 1(c). The transverse electric (TE) mode is excited from a nanowire Si waveguide at a wavelength of 1520 nm. We can observe that the electrical field is penetrating well into the PhC, which indicates that the mode is a rather gap-guided mode [19].

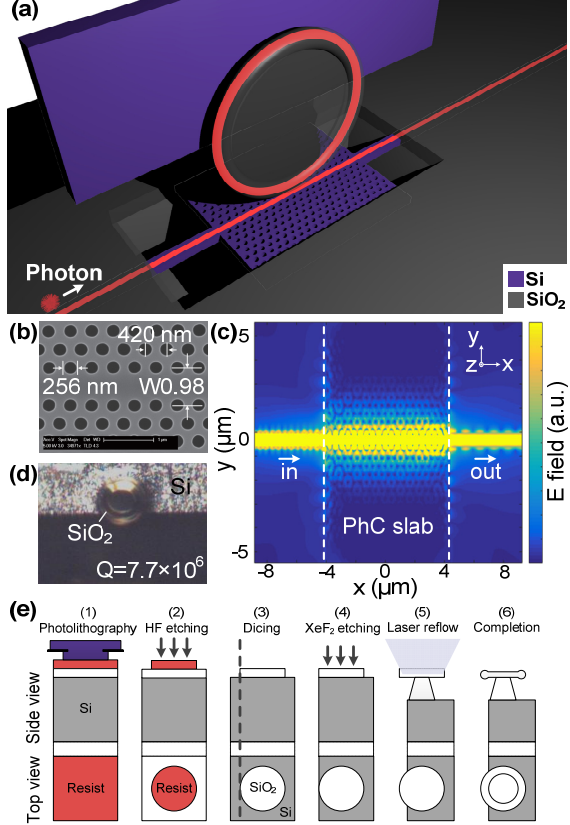

Fig. 1. (a) Schematic illustration of the coupling system. An edge silica toroid microresonator is evanescently coupled with a PhC-WG. The silica cladding in the center is removed. (b) Scanning electron microscope image of the fabricated W0.98 PhC-WG. (c) Propagation mode of the W0.98 PhC-WG at 1520 nm calculated with the 3D finite-difference time-domain (FDTD) method. (d) Microscope image of the edge silica toroid microresonator. The major and minor diameters of the toroid microresonator are 60 and 10 µm, respectively. (e) Schematic illustration of the fabrication of an edge-type silica toroid microresonator.

Figure 1(d) shows a microscope image of the edge-type silica toroid microresonator whose intrinsic (unloaded) $Q$ ($Q_i$) is $Q_i = 7.7 \times 10^6$, which was measured by using a standard tapered fiber setup. This edge silica toroid microresonator is fabricated as shown in Fig. 1(e), where the process is divided into five parts; (1) photolithography, (2) $SiO_2$ etching, (3) Si dicing, (4) $XeF_2$ dry etching, and (5) laser reflow.

A schematic diagram of the optical measurement setup is shown in Fig. 2(a). The transmittance of the PhC waveguide is acquired by a power meter while the wavelength of the tunable laser diode is scanned to obtain transmission spectra. The gap distance $d_g$ between the PhC waveguide and the toroid microresonator is changed with a piezo-electric actuator.

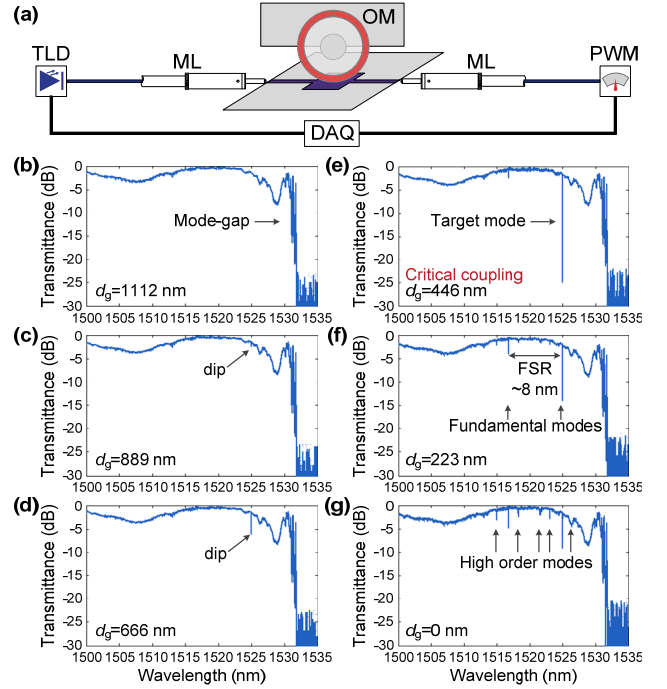

Fig. 2. (a) Experimental setup for coupling between a silica toroid microresonator and a W0.98 PhC waveguide. TLD: Tunable laser diode, ML: Micro-lens, OM: Optical microresonator, PWM: Power meter, DAQ: Data acquisition. (b)-(g) Spectra measured with different gap distances $d_g$. We set $d_g$ = 0 nm when the WGM microresonator touches the PhC waveguide surface, where the absolute $d_g$ value is slightly arbitrary due to experimental limitations (i.e. the toroid microresonator appears to be touching the PhC waveguide surface already when $d_g \sim 100$ nm).

The measured transmittance spectra are shown in Fig. 2(b)-(g), with different $d_g$ values between the toroid microcavity and the W0.98 PhC waveguide. When the WGM resonator is far from the PhC slab we only observe the transmittance property of the PhC waveguide, where we can see a clear cut-off wavelength (mode-gap) at 1532 nm. As we bring the toroid microresonator close to the surface of the PhC waveguide, we start to observe the WGM resonance at 1524.940 nm [Fig. 2(c)] as a dip. When we further decrease $d_g$, the depth of the resonant dip increases and it reaches its maximum value, which is known as the critical coupling condition. When $d_g$ becomes even smaller, then the dip depth of the 1524.940 nm resonant mode becomes shallower, which indicates that the mode is now in an over-coupled regime. In this regime, we start to observe higher order modes [Fig. 2(g)].

Figure 3(a) summarizes the dip depth as a function of $d_g$ for the modes at a wavelength of 1524.940 nm. We see clear evidence for the achievement of the critical coupling condition at around $d_g$ = 446 nm, where Fig. 3(b) is the transmittance spectrum of the resonance at this position. The measured loaded $Q$ ($Q_L$) is

$Q_L = 1.34 \times 10^6$ and the resonant peak has a depth of 23 dB, which shows that the coupling efficiency is very high at up to 99.5%.

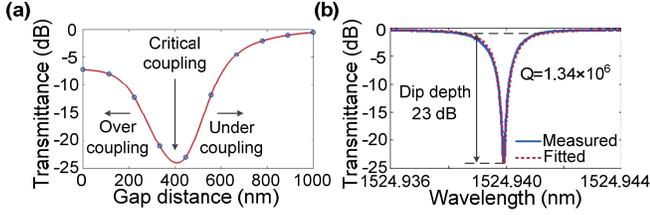

Fig. 3. (a) Relationship between the transmittance of the resonant peak at 1524.940 nm and the gap distance between the PhC waveguide and toroid microresonator. (b) Enlarged view of the resonant peak at critical coupling shown in Fig. 2(e).

It is known that the $Q$s are given by,

$$Q_L^{-1} = \left(Q_i^{-1} + Q_p^{-1}\right) + Q_c^{-1} \quad (1)$$

where, $Q_p$ and $Q_c$ are the parasitic and coupling $Q$s, respectively. Note that $Q_p$ and $Q_c$ are $d_g$ dependent; hence $Q_L$ is also dependent on $d_g$. On the other hand, the transmittance $T$ at the resonant wavelength is given as,

$$T = \left[\frac{\left(Q_i^{-1} + Q_p^{-1}\right) - Q_c^{-1}}{\left(Q_i^{-1} + Q_p^{-1}\right) + Q_c^{-1}}\right]^2 \quad (2)$$

Hence, we obtained $Q_c = 2.68 \times 10^6$ at the critical coupling condition, from the measured $Q_L$ and Eq. (1). The $Q_c$ value is even smaller, namely $1.16 \times 10^6$, when the WGM-PhC WG system is at over-coupling regime at $d_g = 0$ ($Q_p \sim 4.37 \times 10^5$ is used for this calculation). It should be noted that such surprisingly efficient coupling (99.5% at the critical coupling condition) is normally impossible to achieve when the coupling occurs between a resonator made of low-index material with a photonic chip that is made of high-index material. The critical coupling was possible only because we chose a PhC waveguide as a coupler, as we discuss in the following sections.

To study the principle behind the coupling system, we first calculated the dispersion line of an air-bridge type W0.98 PhC waveguide, as shown in Fig. 4(a). Guided modes are introduced in the gap (pink shaded region), and they are classified as even (solid lines) and odd (dashed lines) modes. The blue shaded region is the light cone, which limits the available wavelength range for the application. We usually use the even mode in the PhC waveguide. As shown, the even modes exist within the normalized frequency range from 0.274 to 0.289. Here, the frequency $\omega$ and wavevector $k_x$ are normalized by lattice constant $a$ (420 nm).

Next, we transfer the dispersion line of the even mode in Fig. 4(a) to an $n_{eff}$ map according to the equation $k_x = 2\pi n_{eff}/\lambda$, as shown in Fig. 4(b), where $\lambda$ is the wavelength. The blue line is the $n_{eff}$ of the even mode in the W0.98 PhC waveguide. It shows a trend for the rapid growth of $n_{eff}$ with increases in wavelength. More importantly, the $n_{eff}$ of the even mode crosses the refractive index of silica ($n_{silica} = 1.44$) at a wavelength of around 1525 nm. This means that the silica toroid microresonator and the Si W0.98 PhC-WG mode are perfectly phase matched and a high-coupling coefficient is expected at this wavelength. Indeed, this is the reason for having high coupling at this wavelength in the experiment as shown in Fig. 2(e).

In addition to the appearance of the WGM mode at 1524.940 nm, Fig. 2(e) shows that WGM modes at different wavelengths also couple to the PhC WG when $d_g$ is small. Although the $n_{eff}$ values are not a perfect match, they have reasonably similar values according to Fig. 4(b), which allows these modes to couple as well at a small $d_g$.

The strong coupling between the silica WGMs and the waveguide mode is a unique property of a PhC waveguide, and it is impossible to achieve with a simple Si nanowire waveguide structure whose $n_{eff}$ is shown by the red line in Fig. 4(b). The $n_{eff}$ values are simply too far from that of the WGM resonator. Intuitively, the electrical field of the gap-guided propagating mode in the PhC waveguide penetrates well into the PhC region (Fig. 1(c)) where air holes are present, so $n_{Si} > n_{eff} \sim n_{silica}$ is achieved.

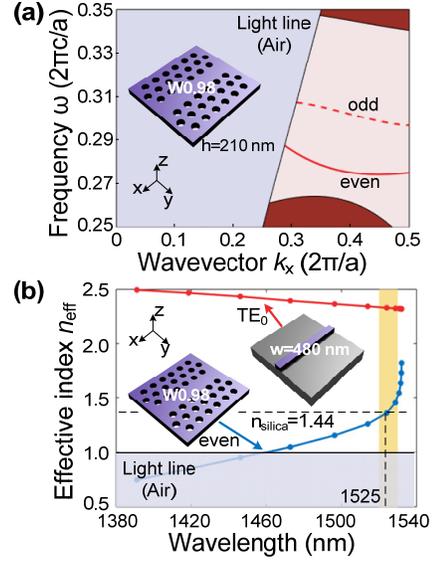

Fig. 4. (a) Computed guided modes of a W0.98 PhC waveguide. (b) Effective indices $n_{eff}$ as a function of wavelength for the fundamental transverse electric (TE$_0$) mode (red line) in a ridge nanowire WG and the even mode (blue line) in a W0.98 PhC waveguide.

Finally, we theoretically investigate and discuss the way in which the group index ($n_g$) of the PhC waveguide will affect the coupling. To obtain information about the coupling between the silica WGM and the Si PhC waveguide, we performed a 3D FDTD calculation for the structure shown in the inset of Fig. 5(a). A straight air-clad fiber with a diameter of 1 μm is assumed as a replacement for a silica WGM resonator. It is placed on the top of a 40.32-μm long PhC waveguide with a gap distance $d_g = 100$ nm. We inject light from the PhC waveguide (from Port 1) and monitor the transmittance (at Port 2). When the light drops towards the thin air-clad fiber (toward Ports 3 and 4), we should be able to observe a transmittance dip. The $n_{eff}$ values for two different PhC waveguides (W0.98 and W1.02) are shown in Fig. 5(a). Since the $n_{eff}$ of the fiber is $n_{fiber} = 1.17$, we expect to obtain efficient coupling at the wavelength indicated by the yellow shading.

The 3D FDTD calculation results for those two PhC waveguides are shown in Fig. 5(b) and (c). Indeed, a strong dip is observed at the wavelength where the phase matching between the PhC and fiber waveguides is satisfied. This clearly shows the importance of the $n_{eff}$ matching. In addition to this, we monitor the dip depth, which provides us with information on the coupling strength. A clear difference can be seen between these two PhC waveguides. One with a dip depth of 13.0 dB but the other having a dip depth of only 10.5 dB. We attribute this difference to the different $n_g$ values at these wavelengths.

It is known that a PhC waveguide exhibits an extremely large $n_g$ when the wavelength is close to the mode gap [19]. Since the wavelength location with the W0.98 waveguide is closer to the mode gap, $n_g$ is larger ($n_g$ = 7.6 @ 1504 nm) than that for W1.02 ($n_g$ = 6.4 @ 1518 nm), and this must contribute to the different coupling strength. When $n_g$ is large, the light in the PhC waveguide can interact with the silica fiber for a much longer time, which allows us to realize a stronger coupling coefficient ($\kappa$) for the same interaction length. Indeed, we calculated the result for different waveguide widths with different $n_g$ values and obtained the transmittance dips and the $\kappa$ values shown in Fig. 5(d). $\kappa$ was calculated based on the dip of resonant peak and the length of the coupling area. This also suggests that a larger $n_g$ will enable stronger coupling. Since a large $n_g$ is another unique property of the PhC waveguide, we can conclude that the PhC waveguide is very attractive as a choice for coupler if we want directly couple a WGM in a Si photonic chip.

In summary, we have demonstrated the efficient coupling of the silica WGM to a high-index Si chip by employing a PhC waveguide. As a result of strong coupling (i.e. a small $Q_c$ of > 1.16×10$^6$), we successfully achieved critical coupling, where the coupling efficiency was 99.5%. This is not usually possible if we use a simple Si wire waveguide structure, due to the large $n_{eff}$ mismatch. In addition to the phase velocity matching, the large $n_g$ (i.e. small group velocity) of the PhC waveguide also contributes to the achievement of efficient coupling. A PhC waveguide allows us to obtain phase index matching ($n_{eff}$ matching), and a large group index ($n_g$) simultaneously, and both contribute to the efficient coupling of the WGM directly to the Si chip. These results provide a robust and efficient method for coupling a low refractive index resonator to a Si platform and will lead to the possibility of benefitting from the unique WGM resonator properties on a Si chip.

**Funding.**
Part of this work was supported by JSPS Kakenhi (#JP16K13702 and JP19H00873), and Strategic Information and Communications R&D Promotion Programme (SCOPE) (#191603001) from the Ministry of Internal Affairs and Communications. The first author is grateful for the financial support provided by the China Scholarship Council (No. 201808320337) and the Postgraduate Research & Practice Innovation Program of Jiangsu Province (No. KYZZ16_0251).

**Disclosures.** The authors declare no conflicts of interest.

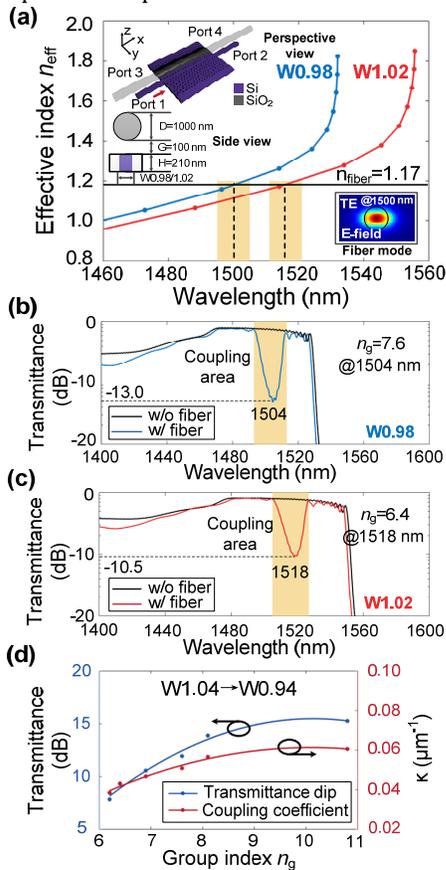

Fig. 5: (a) Effective refractive index as a function of wavelength for two different PhC waveguides (W0.98 and W1.02). Inset (left) is a schematic illustration of the coupling structure for the investigation. An air-clad silica fiber 1 μm in diameter is placed on top of a PhC waveguide with gap distance of $d_g$ = 100 nm. Inset (right) is the cross-section of the electrical field of the TE mode of the fiber waveguide. (b) The calculated transmittance spectrum of a W0.98 PhC waveguide with and without the silica fiber. (c) As (b) but with a different width of W1.02. (d) Relationship between group index $n_g$ and dip, and that between group index $n_g$ and coupling coefficient $\kappa$ for waveguides with different PhCs ranging from W0.94 to W1.04.